# P-TimeSync: A Precise Time Synchronization Simulation with Network Propagation Delays

Wei Dai
Department of Computer Science
Purdue University Northwest
Hammond, Indiana, USA
weidai@pnw.edu

Rui Zhang
Department of Computer Science
Purdue University Northwest
Hammond, Indiana, USA
zhan5097@pnw.edu

Jinwei Liu
Department of Computer and
Information Sciences
Florida A&M University
Tallahassee, Florida, USA
jinwei.liu@famu.edu

*Abstract*—Time serves as the foundation of modern society and will continue to grow in value in the future world. Unlike previous research papers, authors delve into various time sources, ranging from atomic time and GPS time to quartz time. Specifically, we explore the time uncertainty associated with the four major Global Navigation Satellite Systems. In existing time synchronization simulations provide partial usages. However, our research introduces a comprehensive and precise time synchronization simulation named P-TimeSync, leading to a better understanding of time synchronization in distributed environments. It is a state-of-the-art simulation tool for time because (1) it can simulate atomic clocks and quartz clocks with user-defined software clock algorithms, (2) the simulation provides nanosecond-level precision time across different network propagation paths and distances, (3) the tool offers a visualization platform with classic algorithms for distributed time synchronization, such as Cristian's algorithm and Berkeley algorithm. The simulation easily allows for the redefinition of configurations and functions, supporting advanced research and development. The simulation tool could be downloaded via the website: https://github.com/rui5097/purdue_timesync

## I. INTRODUCTION

Precise time is essential for a variety of reasons in everyday life, industry, and scientist.The modern society runs on an accuracy time. However, incorrect time, especially in critical systems or processes, can lead to a range of issues and challenges that can have significant consequences. The use cases of time synchronization have been illustrated in Fig. 1.

Precise time ensures the synchronization of activities, events, and processes in various domains, such as transportation, communication, and industrial production. Without precise timekeeping, it would be challenging to coordinate schedules and operations, resulting in obvious inefficiencies and disruptions. On the other hand, incorrect time can lead to data inconsistencies and errors, potentially compromising the accuracy and integrity of records. This can be particularly problematic in sectors such as finance, healthcare, and research, where precise time stamps are essential for tracking events and activities.

Precise time is fundamental for the functioning of transportation systems, including air, sea, and land navigation. It is crucial for maintaining schedules, ensuring the safety of travelers, and coordinating the movement of vehicles and vessels across various time zones. In addition, many industries are subject to strict regulatory requirements and compliance standards that mandate accurate timekeeping for record-keeping, auditing, and reporting purposes. Incorrect time can lead to compliance violations, legal disputes, and regulatory non-compliance, resulting in financial penalties and reputational damage for organizations.

Inaccuracy time could pose safety risks, potentially leading to accidents, errors in patient care, and disruptions in emergency response systems. Precise timing is critical for ensuring the safety and well-being of individuals, and any inaccuracies in timekeeping can compromise the effectiveness of safety measures and protocols. It is essential for coordinating activities, analyzing timelines, and ensuring the accuracy of security events captured by surveillance cameras, access control systems, and other monitoring devices.

The major contributions of the research have been summarized as follows.

This paper critically examines diverse time sources, atomic time, GPS time, and quartz time. Our focus is particularly on elucidating the time uncertainty inherent in the four major Global Navigation Satellite Systems.

After discussing two time-simulation functions, Eqs. 2-3, that are broadly used in research papers. Surprisingly, the inaccurate time-simulation function, Eqs 3, proposed [1] in 1987 with 1100 citations. We pointed out the major limitations of soft clock functions with mathematics proofs as shown in Eqs. 3-11.

The paper proposes a precise clock simulation with time synchronization function and network propagation delays. To be best of our knowledge, this is a new simulation combining the time function and the propagation delays together.

The remainder of the paper is organized as follows. Section II is the related work. Section III discusses two software clock functions. Section IV introduces the clock simulation with three user scenarios. Section V provide discussion of P-TimeSync simulation. Section VI, and VIII are conclusion, and future work, respectively.

## II. RELATED WORK

### A. Atomic time, GPS time, and Quartz Time

In 1955, The utilization of atomic properties for time measurement originated with the commencement of regular operation of the first cesium beam frequency standard at the National Physical Laboratory in the United Kingdom [2].

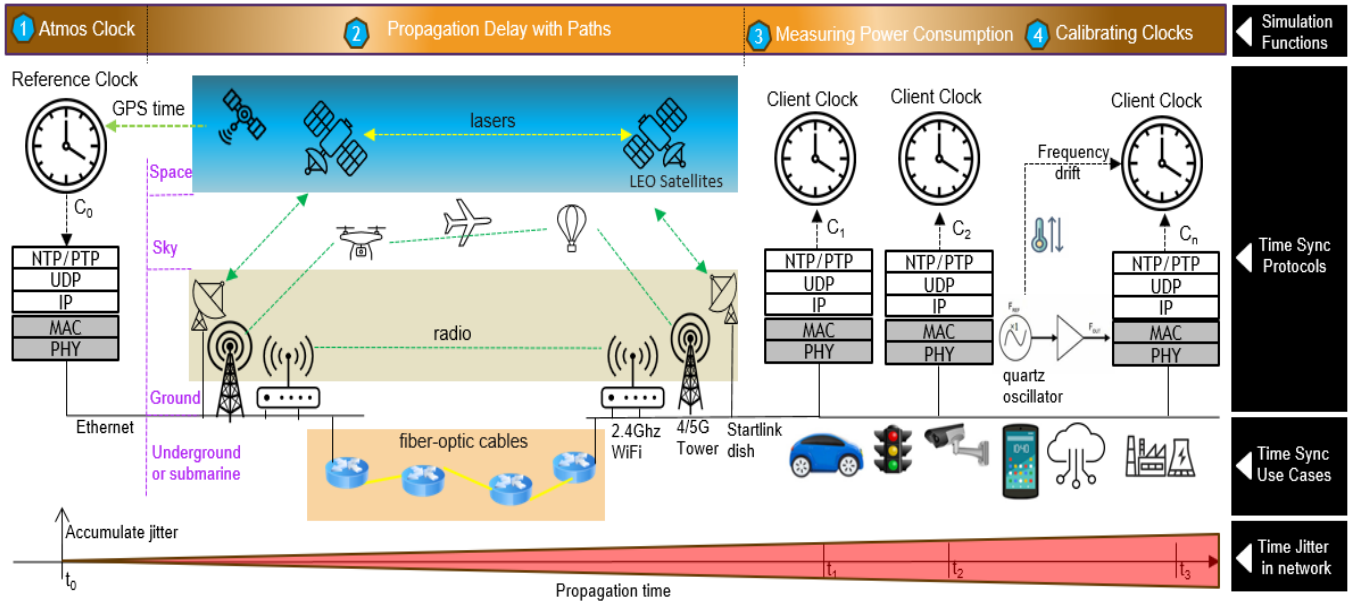

Figure 1. A Data Flowchart of the Clock Synchronization

Now the National Institute of Standards and Technology (NIST) in the United States utilize NIST-F2, a cesium atomic clock, as the primary time and frequency standard. NIST-F2 will not gain or lose a second in at least 300 million years [3].

There are four major Global Navigation Satellite Systems (GNSS) that provide earth position, velocity, and high-accuracy time. GNSS includes the GPS (Global Positioning System) of the United States, the BeiDou (BeiDou Navigation Satellite System) of China, the Global Satellite Navigation System (Galileo) of the European Union, and GLONASS (Global Navigation Satellite System) of Russia. NIST remotely recalibrates GPS time, resulting in within uncertainty of 2 picosecond at one day[4].

TABLE I. GNSS AND TIME UNCERTAINTY

| GNSS, Owner | Uncertainty of Time Transfer (nanoseconds) |
|---|---|
| GPS, USA[5] | [0, 30) |
| BeiDou, China [6][7] | [0, 50) |
| Galileo, EU[5] | [0, 30) |
| GLONASS[5] | [0, 40) |

Note: when comparing uncertainty of time transfer in four GNSS receivers, we only compare one-way communication instead of two-way communications. The BeiDou could achieve [0,20) nanoseconds under two-way communication tests.

In most industry and science, people choose cost-effective **m**iniaturized **r**ubidium **c**locks (MRC) and quartz time. The AR133 series of the AccuBeat company are MRCs with gaining or losing 1 microsecond at 24 hours [8]. The SA.3Xm of Microsemi is gaining or losing 1.5 microsecond per day [9]. However, [10] discovered that frequency stability of MRCs is sharply degraded by at least one order of magnitude when testing these clocks in high-dynamic environments with up to 9m/s$^2$ accelerations. These research results indicate that MRCs also should be frequently recalibrated.

Quartz oscillators are widely used in industry as frequency resources because their price is affordable. [10] pointed out that quartz oscillator of Stanford Research Systems SC10 shows a significant g-sensitivity degrading about two orders of magnitudes in high-dynamic environments. [11] discovered that quartz oscillators are impacted by environment, including temperature, humidity, pressure, acceleration and vibration, magnetic field, electric field, load, and radiation. The authors pointed out that changing temperature significantly shifts frequency.

*B. Time Synalization Protocals*

Both the Network Time Protocol (NTP) and Precision Time Protocol (PTP) are widely accepted in the industry, with NTP being particularly prevalent. PTP stands out as a hardware-based time synchronization technology. The distinctions between NTP and PTP are detailed in Table II.

TABLE II. NTP AND PTP

| Feature | NTP | PTP |
|---|---|---|
| Precision | Millisecond-level | Nanosecond-level |
| Solution | Software-central solutions | Hardware-central solutions with special NICs |
| Cost | Affordable | Expensive |

### III. SOFTWARE CLOCKS

Assume that *t* is the wall clock, denoting ground-truth time or actual time. $C(t)$ is the software clock. Practically speaking, we always observe that $C(t)$ approximately equals to *t*, $C(t) \approx t$. Frequency drift, frequency offset, and random variations could impact the software clock, resulting in time deviation. $\alpha(t)$ is the time deviation between software clock and wall-clock as shown in Eq. (1). In computer science, the software clock is the system time of computer or IoT devices. Note that the time offset, $\alpha(t)$, would gradually increase when observing for a long time.

$$\alpha(t) = C(t) - t \qquad (1)$$

In the mathematical model, software clocks are represented by polynomial regressions, serving as simulations of physical clocks. Two distinct mathematical formulas are employed to describe software clocks: the linear model and the second-order polynomial.

Equation (2) is the linear model.

$$C(t) = \alpha_0 + \beta t + \varepsilon(t) \quad (2)$$

where $\alpha_0$ is the time-offset. $\beta$ is frequency offset. $\varepsilon(t)$ is the random noise. In [12], authors stated that the linear model was called the simple skew model, and $\beta$ was the correlation between wall clock and software clock.

Equation (3) is the second-order polynomial [1][13].

$$C(t) = \alpha_0 + \beta t + \gamma t^2 + \varepsilon(t) \quad (3)$$

where $\alpha_0$ is the time-offset. $\beta$ is frequency offset. $\gamma$ is frequency drift. $\varepsilon(t)$ is the random noise. In [1], authors states that $\gamma$ is not a constant number. The only exception is the cesium clock, in which we set $\gamma$ to zero.

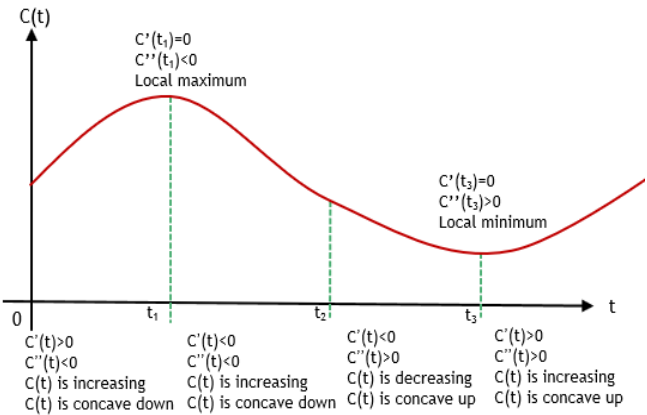

Figure 2. The Two-differentiable Software Clock *C(t)*.

When $C''(t) < 0$ for $t < t_2$, the function C(t) is concave down over the interval $(0, t_2)$. When $C''(t) < 0$ for $t > t_2$, the function c(t) is concave down over the interval $(t_2, +\infty)$. Note that $C'(t) = 0$ has a local maximum at *t* if $C''(t) < 0$. Also, $C'(t) = 0$ has a local minimum at *t* if $C''(t) > 0$. For details, see Fig. 2.

Now, we apply the second-derivative test to determine whether the function exhibits a local maximum or a local minimum.

$$\because C'(t) = \frac{d}{dt}(\alpha_0 + \beta t + \gamma t^2 + \varepsilon) = \beta + 2\gamma t \quad (4)$$

$$C''(t) = \frac{d}{dt}(\beta + 2\gamma t) = 2\gamma \quad (5)$$

$$\therefore C''(t) = \frac{d^2}{dt^2}(\alpha_0 + \beta t + \gamma t^2 + k) = 2\gamma \quad (6)$$

Equations 4 and 6 indicate that $\beta$,frequency offset, and $\gamma$, frequency drift, impact software clocks. If $\gamma > 0$, $C(t)$ has local minimum. Otherwise, C(t) has local maximum. According to Eq. 4, we are able to find the minimal or maxima valuable.

$$\because \beta + 2\gamma t = 0 \quad (7)$$

$$\therefore t = \frac{-\beta}{2\gamma} \quad (8)$$

In [13], authors provided parameters of time simulation: $\beta = 10 \times 10^{-6}$ and $\gamma = -1 \times 10^{-10}$. Then, we calculate $C'(t)$ and $C''(t)$ as shown in Eqs. 9-11.

$$\because C'(t) = \beta + 2\gamma t = 0 \quad (9)$$

$$\therefore t = \frac{-\beta}{2\gamma} = \frac{-10 \times 10^{-6}}{2 \times (-1) \times 10^{-10}} = \frac{10^{-5}}{2 \times 10^{-10}} = 5 \times 10^4 \quad (10)$$

$$C''(t) = 2\gamma = -2 \times 10^{-10} < 0 \quad (11)$$

According to Eqs. 9-11, we know that when t = $5 \times 10^4$ seconds or 13 hours 53 minutes, the time-offset of software clock $C(t)$ is the local maximum. However, in the real world, the time-offset of clock does not automatically reduce after passing a specific time slot. Thus, we demonstrate that if we assign constant numbers to $\beta$ and $\gamma$, the simulation is imperfect.

IV. A PRECISE TIME SYNCHRONIZATION SIMULATION

We create the precise time synchronization simulation (or P-TimeSync for short) based on probability theory, graphic algorithms, and discrete events. This simulation program uses the graph-tool library [14] to implement graph theory algorithms and visualization for networks. The simulation could be run on the Google Colab or Jupyter Notebook. The simulation tool could be downloaded via the website as follows:

https://github.com/rui5097/purdue_timesync

P-TimeSync supports user-defined software clocks, network routers (Wi-Fi routers and regular routers), network bandwidth, and network distances, two distributed time synchronization algorithms, and security of time synchronization.

a) Software Clocks

The simulation supports one software clock function to multiply virtual clocks. Excepting Eqs 1-2, the simulation supports user-defined time functions.

b) Network Routers

It supports Wi-Fi routers and regular routers. In the real world, when the router has failure, the data travels could be changed. So, the simulation supports router failure rates with real-time function flag(t). When flag(t)≡1, the router is active. Otherwise, the router is inactive. Thus, the router statues are described in Eq. 12.

$$tRouter(t) = \sum_{i=1}^{n} Flag(t) \times Router\_delay(i) \quad (12)$$

where *i* is router number at time *t*. The accumulated delay time of routers, $tRouterPath()$, is all delay time of active routers in the travel path. Note that the delay time of active routers are configured via user-defined parameters. Inactive routers will be ignored because their delay time is infinite, ∞.

c) Network Bandwidth

Customers have the flexibility to define the network bandwidth, ranging from Kbps to Gbps, allowing for the simulation of low-bandwidth wireless networks and fiber networks. The network bandwidth directly influences the data transmission time, as illustrated in Eqs. 13-14.

$$TravelBW\_delay_i = \frac{Size-of-Transmission}{network\_bandwidth(i)} \quad (13)$$

$$tTravelData(t) = \sum_{i=1}^{n} TravelBW\_delay(i) \quad (14)$$

d) Network Distance

It supports network distance because network distance and network types impact travel speeds, as illustrated in Eqs 15-16.

$$TravelDista\_delay_i = \frac{distance}{network\_speed(i)} \quad (15)$$

$$tTravelDistance(t) = \sum_{i=1}^{n} TravelDista\_delay(i) \quad (16)$$

e) Distributed Time Synchronization Algorithms: the simulation supports the Berkeley clock synchronization algorithm and Cristian's algorithm.

## V. Discussion

In the real world, data travel paths are unstable due to router situations. In the simulation, we employ Dijkstra's algorithm to address the short path problem when data is transmitted across different routers. It's important to note that the path for sending data and the path for receiving data may differ, as router situations can change over time. In Figs 3-4, we show the visualization results of P-TimeSync simulation. For details, please read the open-source website.

The total travel time includes router delays, $tRouter$, data delay, $tTravelData$, and distance delay, $tTravelDista$.

$$tTravelTotalDelay(t) = \sum_{i=1}^{n}[\,tRouter(i) + tData(i) + tTravelDistance(i)\,] \quad (17)$$

## VI. Conclusion

In this paper, authors discussed two existing software clock algorithms. While the linear model may not capture complex situations, the second-order polynomial model has inherent limitations, as demonstrated through two differentiable approaches. Additionally, we present the P-TimeSync simulation, an open-source Python-based tool. This simulation supports user-defined software clock functions, facilitates the determination of the shortest data travel paths and calculates network propagation delays.

## VII. Future work

There are three potential approaches for improving the accuracy of time synchronization. Firstly, machine learning algorithms could identify patterns in network delay time, thereby enhancing the accuracy of distributed time sync algorithms. Secondly, enhancing time sync security is crucial for future algorithms, especially as precision timing becomes increasingly vital for financial programming and distribution systems. Lastly, the exploration of IoT energy in the context of time synchronization should be considered in future research.

## Acknowledgment

We express our gratitude to Mr. Sirish Chejerla for his involvement during the initial stage of the research project in October 2023.

Figure 3. The screenshot of P-TimeSync Simulation with Wi-Fi routers and fiber network

Figure 4. The screenshot of P-TimeSync Simulation with fiber networks and Startlink communications

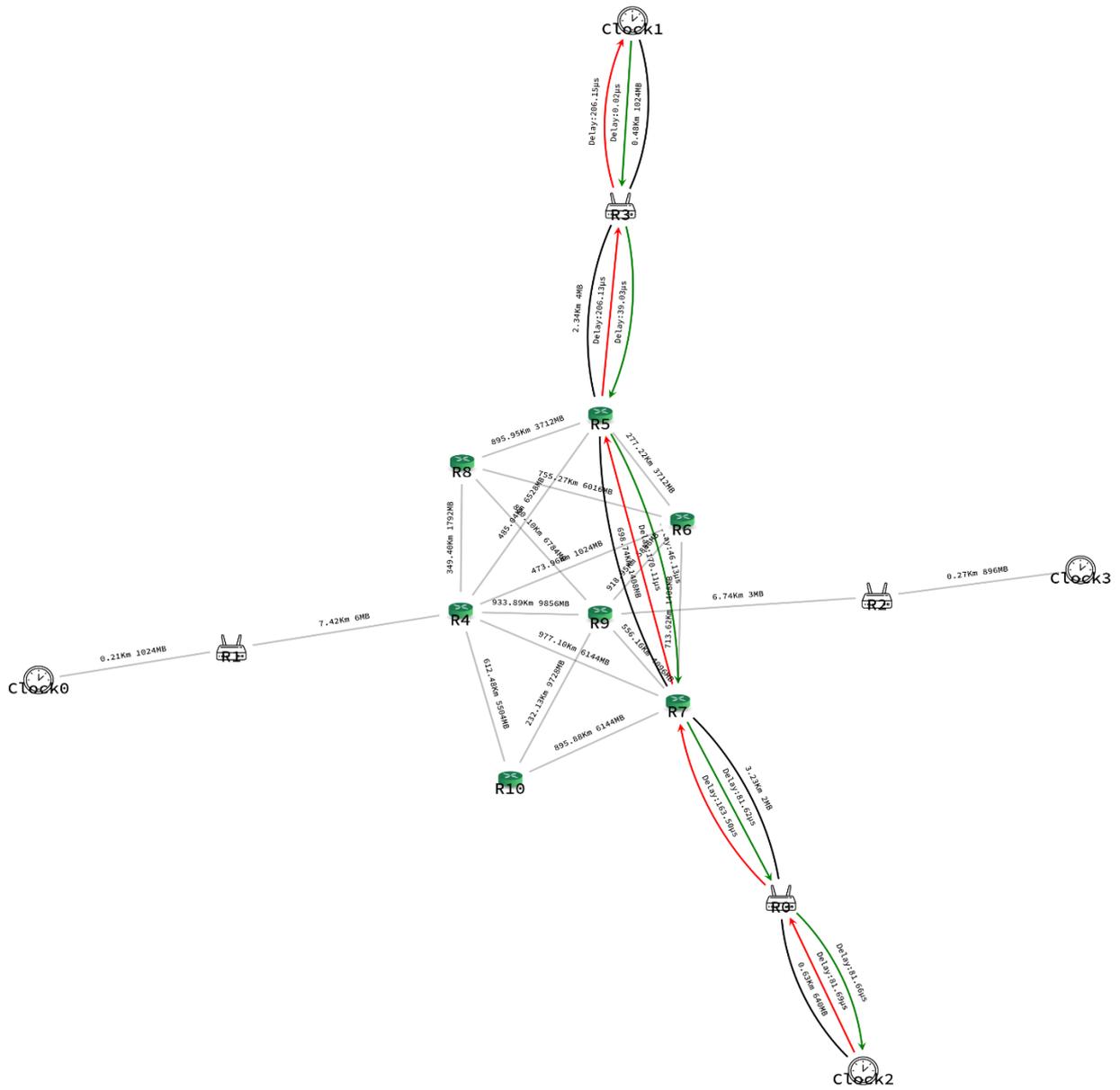
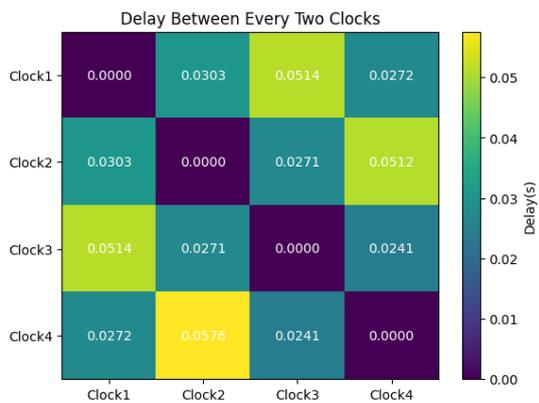
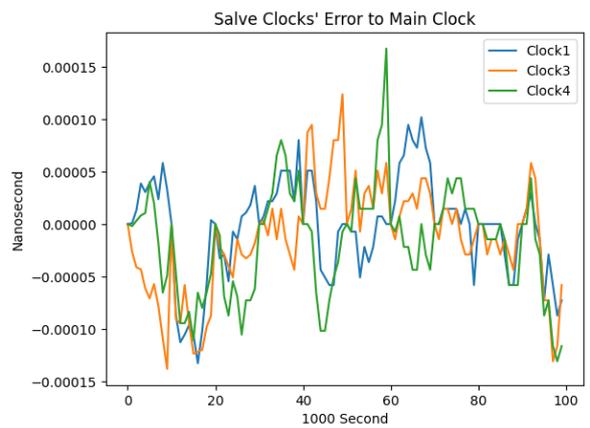

Figure 5. The screenshot of P-TimeSync Simulation with the shortest path on network routers.